\documentclass[3p,sort&compress]{elsarticle}

\usepackage{lineno,hyperref}
\usepackage{amssymb}
\usepackage{amsmath}

\begin{document}

\begin{frontmatter}

\title{ Which-Way Information in Double-Slit and COW Experiments \\ with  Unstable Particles}

\author[Wabash,Purdue]{D.E.~Krause\corref{cor1}}  \ead{kraused@wabash.edu}

\address[Wabash]{Physics Department, Wabash College, Crawfordsville, IN 47933, USA}
\address[Purdue]{Department of Physics and Astronomy, Purdue University, West Lafayette, IN 47907, USA}

\author[Purdue]{E.~Fischbach}

\author[Avon]{Z.J.~Rohrbach}
\address[Avon]{Avon High School, 7575 East 150 South,
Avon, IN 46123, USA}

\cortext[cor1]{Corresponding author}

\begin{abstract}
One  might expect that a quantum undecayed unstable particle (QUUP) should behave in the same manner as an identical, albeit stable, particle,  but it turns out that this is not always true.  We show explicitly that using QUUPs in the double-slit and Colella-Overhauser-Werner (COW) experiments leads to {\em a priori} which-way information that creates a loss of interference contrast when compared to the same experiments performed using stable particles.   In both of these cases,  {\em a priori}  path predictability ${\cal P}$ is related to the interference visibility ${\cal V}$ by the duality relation ${\cal P}^{2} + {\cal V}^{2} = 1$.

\end{abstract}


\end{frontmatter}


\section{Introduction}

Quantum mechanics has been part of physics for nearly ninety years and is used to understand and develop much of our everyday technology.  Yet, while its mathematical application to physical problems is undisputed, the same cannot be said for the conceptual view of the microscopic world that quantum mechanics provides.   Serious issues arose early with the wave-particle duality of the electromagnetic field introduced by Planck and Einstein, which was extended to matter waves by de Broglie.  Depending on the nature of the experiment, these systems exhibited wave or particle-like behavior.

Probably the  system most  commonly used to demonstrate the wave-particle duality is the double-slit experiment \cite{Feynman}.  Under appropriate conditions, the  probability that a particle will strike a screen after passing through two slits exhibits interference in the same form as the intensity of a classical light wave passing through similar slits.  In the quantum case, the wave amplitude leading to  interference  is a  complex vector in an abstract Hilbert space rather than a physical electric field or pressure wave amplitude which can be measured (in principle).  Furthermore, as Feynman demonstrated dramatically \cite{Feynman}, the amount of interference one observes for quantum particles depends on whether one looks to see which slit the particle pass through.  If one doesn't place a detector to observe the slit passage, one sees the full interference pattern, but if the detector is present, the interference pattern disappears.  Thus, the amount of quantum interference one observes depends on the amount of  ``which-slit'' information one obtains.  However, Feynman only considered two extreme cases involving which-way information---having no knowledge or perfect knowledge.  In practice, one only has partial information over which path a particle may have taken, and this case was first investigated for the double-slit experiment by Wootters and Zurek \cite{Wootters Zurek}.  Subsequently, a number of  authors \cite{Glauber,Greenberger Yasin,Jaeger,Englert, Schwindt,Durr AJP} have considered the effect of having partial which-way information  on interference in other two path systems, developing a simple mathematical relationship which we will now consider.

To characterize the amount of interference observed, one usually uses the visibility ${\cal V}$ which measures the contrast of the interference maxima and minima,
\begin{equation}
{\cal V} = \frac{I_{\rm max} - I_{\rm min}}{I_{\rm max} + I_{\rm min}},
\label{V}
\end{equation}
where $I_{\rm max}$ and $I_{\rm min}$ are  adjoining maximum and minimum intensities, respectively, that are observed by a detector as some parameter is varied.

To quantify which-way information for a two-path system, we will follow Englert \cite{Englert} and distinguish two types of quantities.  The thought experiment considered by Feynman would involve  path distinguishability ${\cal D}$, a parameter characterizing {\em a posteriori} which-way information, information gained during the experiment.  In Feynman's case, this would be due to an interaction coupling the particles to the detector.  Alternatively, an unstable particle can give away its position by decaying.  In the case of an excited atom emitting a photon, or a large molecule emitting thermal black body radiation, the particle becomes entangled with the decay products, in which case a loss of coherence and interference results.  One can show the ${\cal D}$ and ${\cal V}$ satisfy the duality relation \cite{Jaeger,Englert,Schwindt,Durr AJP}
\begin{equation}
{\cal D}^{2} + {\cal V}^{2} \leq 1,
\end{equation}
where the equality occurs when the system is in a coherent state.
For Feynman's extreme cases, perfect path detection (${\cal D} = 1$) results in complete loss of interference contrast (${\cal V} = 0$), while no path detection (${\cal D} = 0$) leads to maximum contrast in the interference pattern (${\cal V} = 1$).

  In this paper, we are focusing our attention on quantum {\em undecayed} unstable particles (QUUPs) that do not decay while in the apparatus, in which case the path distinguishability vanishes.  We use the redundant term {\em undecayed} to distinguish an atom in an excited state (a QUUP)   from the same atom in the ground state after it has decayed, either of which can contribute to an interference pattern.
  One might then expect that the interference pattern would be the same for QUUPs as for stable particles in the same experiment, but we will show that in the double-slit and Colella-Overhauser-Werner (COW) \cite{COW} experiments that this is not the case.  In these cases, the instability creates an asymmetry between the probability that the QUUP took one path over the other which provides which-way information available {\em before} the experiment is performed.  
  
  Following Englert, we will quantify this {\em a priori} which-way information using path predictability \cite{Greenberger Yasin,Englert}, which we will write as \cite{Glauber,Greenberger Yasin,Jaeger,Englert,Schwindt,Durr AJP}
\begin{equation}
{\cal P} \equiv \left|\frac{P_{1} - P_{2}}{P_{1} + P_{2}}\right|,
\label{P}
\end{equation}
where $P_{i}$  is the probability that a particle  will reach the detector via  path $i$.   This definition accommodates situations where  $P_{1} + P_{2} < 1$, i.e., when some of the particles that enter the apparatus fail to reach either detector.   This may occur when one uses unstable particles or when an absorbing medium is placed in the particle path as in the experiments Summhammer, et al. \cite{Summhammer}. One can show that ${\cal P}$ and ${\cal V}$ satisfy \cite{Glauber,Greenberger Yasin,Englert,Jaeger,Durr AJP}
\begin{equation}
{\cal P}^{2} + {\cal V}^{2} \leq 1.
\label{P V}
\end{equation}

The purpose of this paper is to investigate the use of unstable particles in  double-slit and COW-type experiments \cite{COW}, demonstrating that the resulting interference satisfies the duality relation Eq.~(\ref{P V}).
 (A third case, involving  a Mach-Zehnder-like atom interferometer using excited atoms with an appropriately tuned cavity, has been considered elsewhere \cite{Krause PLA}.)  In the process we will develop a simple formalism that will allow one to calculate interference effects in a stationary-beam experiment for unstable particles  acted upon by any slowly varying  potential $V(\mathbf{r})$.    In the Appendix we show how the stationary-beam results for the double-slit experiment are consistent with previous work that involved using a time-dependent approach.

\section{Formalism}

We begin by introducing the formalism used to calculate the probability amplitudes for QUUPs, generalizing earlier work involving freely propagating QUUPs \cite{Krause PLA}.  To reveal the behavior of unstable particles in quantum interference at lowest order, a simple phenomenological approach is all that is needed.  We modify the usual  non-relativistic Schr\"{o}dinger wave equation describing an unstable particle of mass $m$  experiencing a time-independent potential $V(\mathbf{r})$ by adding an imaginary constant term to the particle's rest energy proportional to its decay rate $\Gamma$:
\begin{equation}
i\hbar\frac{\partial\Psi(\mathbf{r},t)}{\partial t} = \left(H_{0} + V\right)\Psi(\mathbf{r},t),
\label{S E}
\end{equation}
where the Hamiltonian describing a free non-relativistic unstable particle is given %
\begin{equation}
H_{0} = mc^{2} - i\frac{\hbar\Gamma}{2} - \frac{\hbar^{2}}{2m}\nabla^{2}.
\label{H0}
\end{equation}

Since we are interested in determining the probability that a particle will reach a detector irrespective of time, we will consider only stationary beam experiments here.  (In the Appendix we will discuss the double-slit experiment using a time-dependent approach which yields the same answers.)   This means that we can follow Greenberger and Overhauser \cite{G O} and search for  solutions of Eq.~(\ref{S E}) of the form 
\begin{equation}
\Psi_{E}(\mathbf{r},t) = \psi_{E}(\mathbf{r})e^{-iEt/\hbar},
\label{Psi}
\end{equation}
 where $E$ is a real energy.  Substituting Eq.~(\ref{Psi}) into Eq.~(\ref{S E}) gives the time-independent equation
\begin{equation}
\left(H_{0} + V\right)\psi_{E}(\mathbf{r}) = E\psi_{E}(\mathbf{r}).
\label{HVpsi}
\end{equation}

To solve Eq.~(\ref{HVpsi}), we will use a WKB-like approximation which first requires finding the solutions to the unperturbed Schr\"{o}dinger equation.
Let $ \psi_{E,0}(\mathbf{r})$ be the solution of the free particle time-independent Schr\"{o}dinger equation
\begin{equation}
 \hat{H}_{0}\psi_{E,0}(\mathbf{r}) = E_{0}\psi_{E,0}(\mathbf{r}),
\label{H0 psi}
\end{equation}
which can be rewritten as
\begin{equation}
\nabla^{2}\psi_{E,0}(\mathbf{r}) = -\tilde{\mathbf{k}}_{0}^{2}\psi_{E,0}(\mathbf{r}),
\label{k psi0}
\end{equation}
where the magnitude of the complex wave vector $\tilde{\mathbf{k}}_{0}$ is determined by
\begin{equation}
|\tilde{\mathbf{k}}_{0}|^{2} = \tilde{k}_{0}^{2} \equiv (k_{0} + i \kappa_{0})^{2} =  \frac{2m}{\hbar^{2}}\left(E_{0} - mc^{2} + i\frac{\hbar\Gamma }{2}\right) \simeq \frac{p_{0}^{2}}{\hbar} + i\frac{m\Gamma }{\hbar}.
\label{k}
\end{equation}
Here $p_{0}$ is the particle's momentum,  $k_{0}$ and $\kappa_{0}$ are the real and imaginary parts of $\tilde{k}_{0}$, and the non-relativistic limit of the kinetic energy $E_{0} - mc^{2} \simeq p_{0}^{2}/2m$ was used.  Since  $E_{0}$ is  assumed to be real, Eq.~(\ref{k}) can be used to show that
\begin{eqnarray}
k_{0}^{2} & = &  \frac{p_{0}^{2}}{\hbar^{2}} + \kappa_{0}^{2},\\
\kappa_{0} & = & \frac{m\Gamma }{2\hbar k_{0}}.
\end{eqnarray}
The solutions of Eq.~(\ref{k psi0}) can be written as
\begin{equation}
\psi_{E,0}(\mathbf{r}) = A_{0}e^{i\tilde{\mathbf{k}}\cdot\mathbf{r}},
\end{equation}
where $A_{0}$ is a constant. In order for these solutions to have a well-defined wavelength (i.e., momentum),  $\kappa_{0} \ll k_{0}$, which implies the hierarchy of energy scales $mc^{2} \gg p_{0}^{2}/2m \gg \hbar\Gamma $, leading to
\begin{eqnarray}
k_{0}   & \simeq &    {p_{0}/\hbar} = 1/\lambda_{0}, \\
 \kappa_{0} & \simeq &  m\Gamma/2p_{0} \equiv 1/2\ell_{0}.
 \end{eqnarray}
 Here $\lambda_{0}$ is the free particle de Broglie wavelength, and $\ell_{0}$ is the average distance an unstable particle with momentum $p_{0}$ and decay rate $\Gamma$ travels before decaying. With these approximations, the general solution to the unperturbed time-independent wave equation, Eq.~(\ref{k psi0}), can be written as
\begin{equation}
\psi_{E,0}(\mathbf{r}) \simeq A_{0}\exp\left[{i\mathbf{k}_{0}\cdot\mathbf{r}}\left(1 + i\frac{1}{2k_{0}\ell_{0}}\right)\right].
\label{E0 psi}
\end{equation}

Now to solve the Schr\"{o}dinger equation with a potential, Eq.~(\ref{HVpsi}), we follow Ref.~\cite{G O} and search for solutions of the form 
\begin{equation}
\psi_{E}(\mathbf{r}) = \psi_{E,0}(\mathbf{r})\chi(\mathbf{r}),
\label{psi}
\end{equation}
where $\chi(\mathbf{r})$ is much more slowly varying in position than $\psi_{E,0}(\mathbf{r})$.  Substituting Eq.~(\ref{psi}) into the left side of Eq.~(\ref{HVpsi})  and assuming $|{\nabla}\chi(\mathbf{r})| \ll k_{0}\chi(\mathbf{r})$ leads to
\begin{equation}
\frac{1}{\chi(\mathbf{r})}\frac{d\chi(\mathbf{r})}{ds}  \simeq  i\frac{m}{\hbar^{2}k_{0}}\left(1 - i\frac{m\Gamma}{2\hbar k_{0}^{2}}\right) [E - E_{0} - V(\mathbf{r})] , 
\end{equation}
where $ds$ is the infinitesimal distance traveled along the direction $\mathbf{k}_{0}$.
Integrating over the {\em unperturbed} particle path  gives
\begin{equation}
\chi(\mathbf{r}_{f}) = \chi(\mathbf{r}_{0})\exp\left[-i\frac{m}{\hbar^{2}k_{0}}\left(1 - i\frac{m\Gamma}{2\hbar k_{0}^{2}}\right) \int^{\mathbf{r}_{f}}_{\mathbf{r}_{0}}V(\mathbf{r})\,ds\right],
\label{chi 1}
\end{equation}
where we will only consider situations where  $E = E_{0}$.  
The total time-independent wave function  can then be written as
\begin{eqnarray}
\psi_{E}({\mathbf r}_{f}) & \simeq & A_{0}\chi(\mathbf{r}_{0})e^{ip_{0}s/\hbar}e^{-s/2\ell_{0}}
\nonumber \\
&& \mbox{} \times \exp\left\{-i\frac{m}{\hbar p_{0}}\left[1- i\left(\frac{\lambda_{0}}{2\ell_{0}}\right)\right]\int^{\mathbf{r}_{f}}_{\mathbf{r}_{0}}V(\mathbf{r})\,ds\right\}, \nonumber \\
&&
\label{final Psi E}
\end{eqnarray}
where $s$ is the total distance traveled by the QUUP along its path $\mathbf{r}_{0} \rightarrow \mathbf{r}_{f}$.  Eq.~(\ref{final Psi E})  is identical to the corresponding result for stable particles \cite{G O} after replacing the real free particle wave vector $\mathbf{k}_{0}$ in that result with the complex unstable  particle wave vector $\tilde{\mathbf{k}}_{0}$.

Finally, if we define the potential-dependent complex phase for the path $\vec{r}_{0} \rightarrow \vec{r}_{f}$,
\begin{equation}
\tilde{\phi}_{0f} \equiv -\frac{m}{\hbar p_{0}}\left[1- i\left(\frac{\lambda_{0}}{2\ell_{0}}\right)\right]\int^{\mathbf{r}_{f}}_{\mathbf{r}_{0}}V(\mathbf{r})\,ds,
\label{complex phase}
\end{equation}
then Eq.~(\ref{final Psi E}) simplifies to 
\begin{equation}
\psi_{E}({\mathbf r}_{f})  \simeq A_{0}\chi(\mathbf{r}_{0})e^{ips/\hbar}e^{-s/2\ell_{0}}e^{i\tilde{\phi}_{0f}}.
\end{equation}

\section{Double-Slit Experiment with Unstable Particles}
\label{Double-Slit section}

\begin{figure}[tbp]
\begin{center}
\includegraphics[height=1.3in]{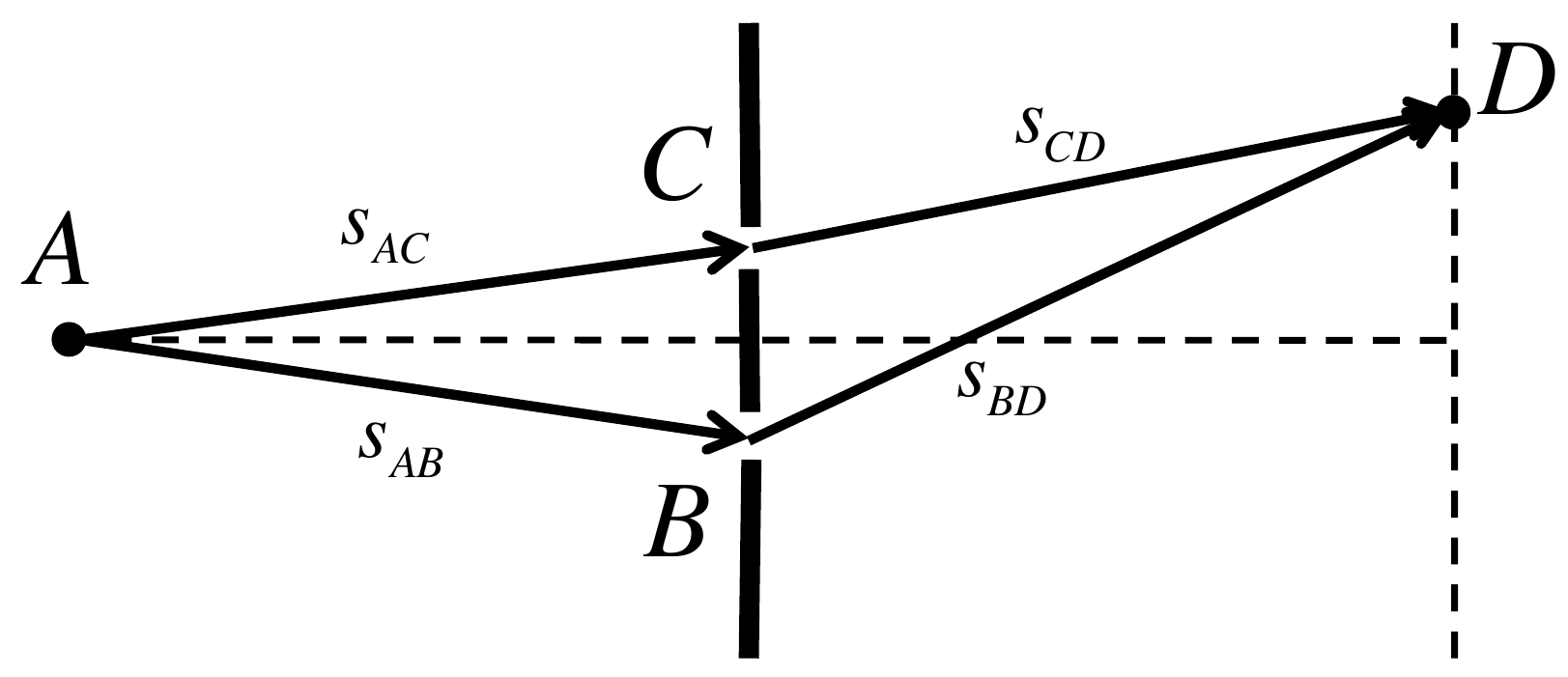}
\caption{Setup for a double-slit experiment.}
\label{Double Slit Figure}
\end{center}
\end{figure}

As a first application of this formalism, consider the setup for a double-slit experiment shown in Fig.~\ref{Double Slit Figure}.
Let $A$ represent a source of QUUPs which can then reach $D$ by passing through either slits $B$ or $C$.  For convenience, the geometry is such that the path lengths between $A$ and $B$ and $A$ and $C$ are the same: $s_{AB} = s_{AC}$.  

\begin{figure}[tbp]
\begin{center}
\includegraphics[width=90mm]{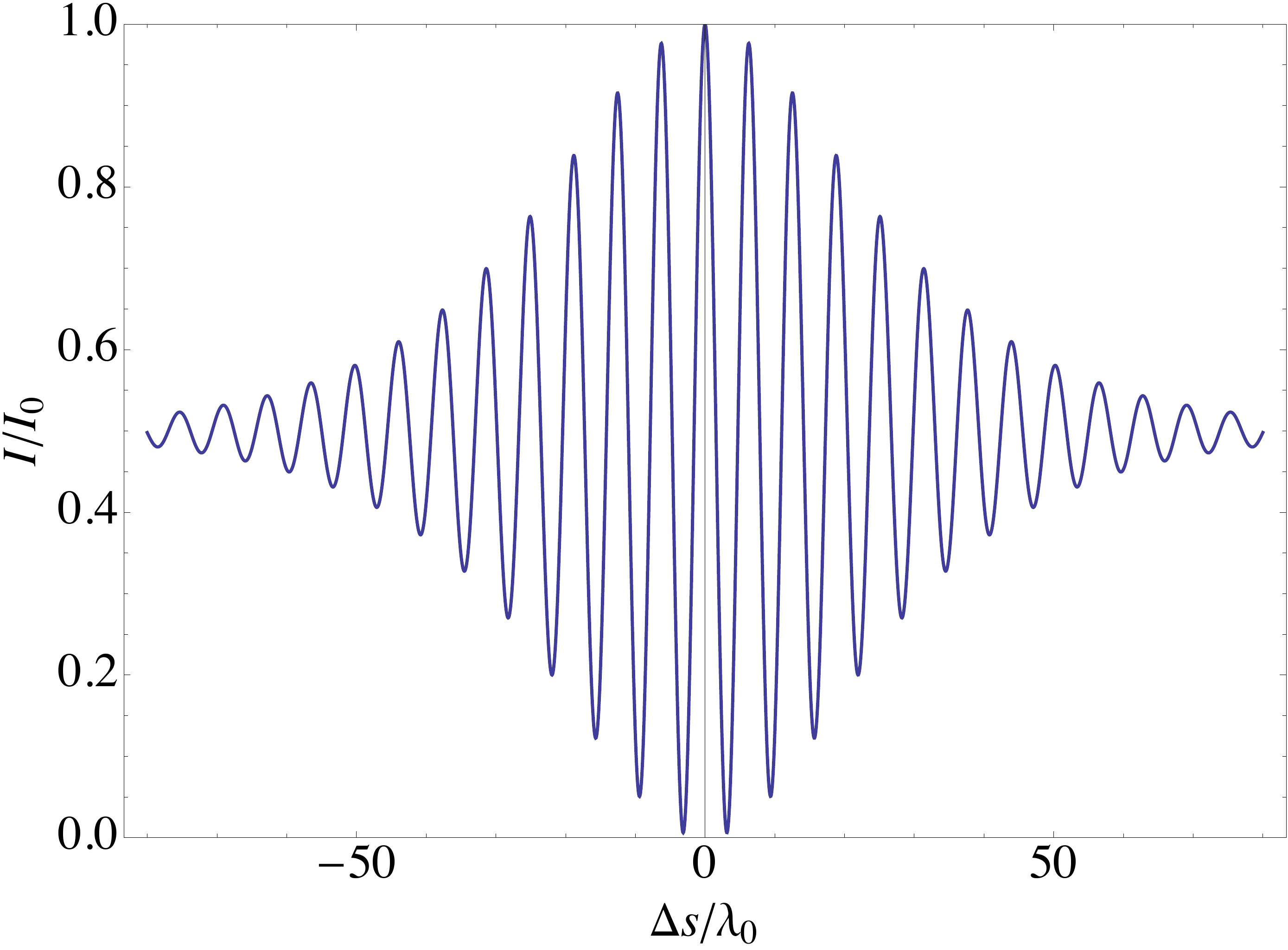}
\caption{Intensity plot for a double-slit experiment obtained from Eq.~(\ref{I DS}) using $\ell_{0} = 10\lambda_{0}$ illustrating the loss of interference contrast for unstable particles as $\Delta s$ increases.}
\label{Double Slit Intensity Figure}
\end{center}
\end{figure}

To determine the probability that the particle will be detected at $D$, we use  Eq.~(\ref{final Psi E}) setting $V(\mathbf{r}) = 0$, giving the free particle amplitude
\begin{equation}
\psi_{E,0}({\mathbf r}) = A_{0}\chi(\mathbf{r}_{0})e^{ip_{0}s/\hbar}e^{-s/2\ell_{0}}.
\label{free Psi E}
\end{equation}
Using Eq.~(\ref{free Psi E}), the probability that the particle will be detected at $D$ after leaving $A$ is obtained from adding the amplitudes for the two paths $ABD$ and $ACD$, leading to:
\begin{eqnarray}
P(D) & = & P_{0}\left[e^{-s_{BD}/\ell_{0}} + e^{-s_{CD}/\ell_{0}} \right.
\nonumber \\
&& \mbox{}\left.
+ 2e^{-(s_{BD}+s_{CD})/2\ell_{0}}\cos\left(\frac{p_{0}\Delta s}{\hbar} \right)\right],
\label{P D}
\end{eqnarray}
where $P_{0}$ is a constant, and $\Delta s = s_{BD} - s_{CD}$ is the path length difference.   The intensity pattern observed on the detection plane results only from those particles that reach it without decaying.  This can be obtained by dividing Eq.~(\ref{P D}) by the sum of the probabilities that the particle traveled the paths $ABD$ and $ACD$ separately, $P_{ABD}$ and $P_{ACD}$,  giving
\begin{equation}
I_{\rm DS}(D) =\frac{I_{0}}{2}\left[1 + {\rm sech}\left(\frac{\Delta s}{2\ell_0} \right) \cos\left(\frac{\Delta s}{\lambda_{0}} \right)\right], 
\label{I DS}
\end{equation}
where $I_{0}$ is the QUUP intensity  when $\Delta s = 0$.  
 An example graph of the intensity obtained from Eq.~(\ref{I DS}) for $\ell_{0} = 10\lambda_{0}$ is shown in Fig.~\ref{Double Slit Intensity Figure}. 
Using Eq.~(\ref{V}), we see that the  visibility for the double-slit interference pattern is given by
\begin{equation}
{\cal V}_{\rm DS} = {\rm sech}\left(\frac{\Delta s}{2\ell_{0}}\right),
\label{V DS}
\end{equation}
which reduces to unity for stable particles  ($\ell_{0} \rightarrow \infty$).

It is now straightforward to show that these results satisfy the relation Eq.~(\ref{P V}).  If $P_{ABD}$ and $P_{ACD}$ are probabilities the particle took paths $ABD$ and $ACD$, from Eq.~(\ref{P}),  the predictability ${\cal P}_{DS}$ for the double-slit experiment is given by
\begin{eqnarray}
{\cal P}_{\rm DS} 
& = & \left|\frac{P_{ACD} - P_{ABD}}{P_{ACD} + P_{ABD}}\right|, \nonumber \\
& = & \left|\frac{e^{-s_{CD}/\ell_{0}} - e^{-s_{BD}/\ell_{0}}}{e^{-s_{CD}/\ell_{0}} + e^{-s_{BD}/\ell_{0}}}\right|,
\nonumber \\
& = & \tanh\left(\frac{|\Delta s|}{2\ell_{0}}\right).
\label{P DS}
\end{eqnarray}
If the particles are stable, or the paths have equal lengths, ${\cal P} = 0$, as expected. It is now clear from the identity $\mbox{sech}^{2}x + \tanh^{2} x = 1$ that Eqs.~(\ref{V DS}) and (\ref{P DS}) satisfy Eq.~(\ref{P V}), and that  they have the same form as the unified description suggested by Bramon, et al. \cite{Bramon}. Using QUUPs in the double-slit experiment thus gives  additional information on which slit the unstable particle took to reach the screen compared to using stable particles.  The QUUP is more likely to have come from the closer slit since it has a greater chance of surviving undecayed, and the effect grows as the path difference increases.  

 It is important  to compare these results to previous investigations of the interference pattern of a double-slit experiment using unstable particles.   In Refs.~\cite{Sleator,Facchi,Takagi}, excited 2-level atoms were used to study the effect that the decay of an unstable particle has on the observed interference pattern.  They found that the visibility of the interference pattern of the decayed atoms which reach the screen decreased significantly when the wavelength of the emitted photon $\lambda_{\rm ph}$ was less than the slit separation $d$, which is when the photon has enough resolution to distinguish which slit the atom passed through. They also found that the time-dependent pattern of the undecayed atoms was the same as for stable particles except for an overall time-dependence $e^{-\Gamma t}$.  However, as we show in the Appendix,  if one treats the undecayed atoms as wave packets with additional time dependence $e^{-\Gamma t/2}$ as in Refs.~\cite{Sleator,Facchi,Takagi}, and then integrates  the intensity arriving at the screen over time to obtain the total probability that the atom arrives undecayed at the detector irrespective of time, one also obtains Eq.~(\ref{P D}) in the limit of infinitely long longitudinal  coherence length (as assumed here).
 
 \section{COW Experiments with Unstable Particles}
 
\begin{figure}[tbp]
\begin{center}
\includegraphics[height=1.7in]{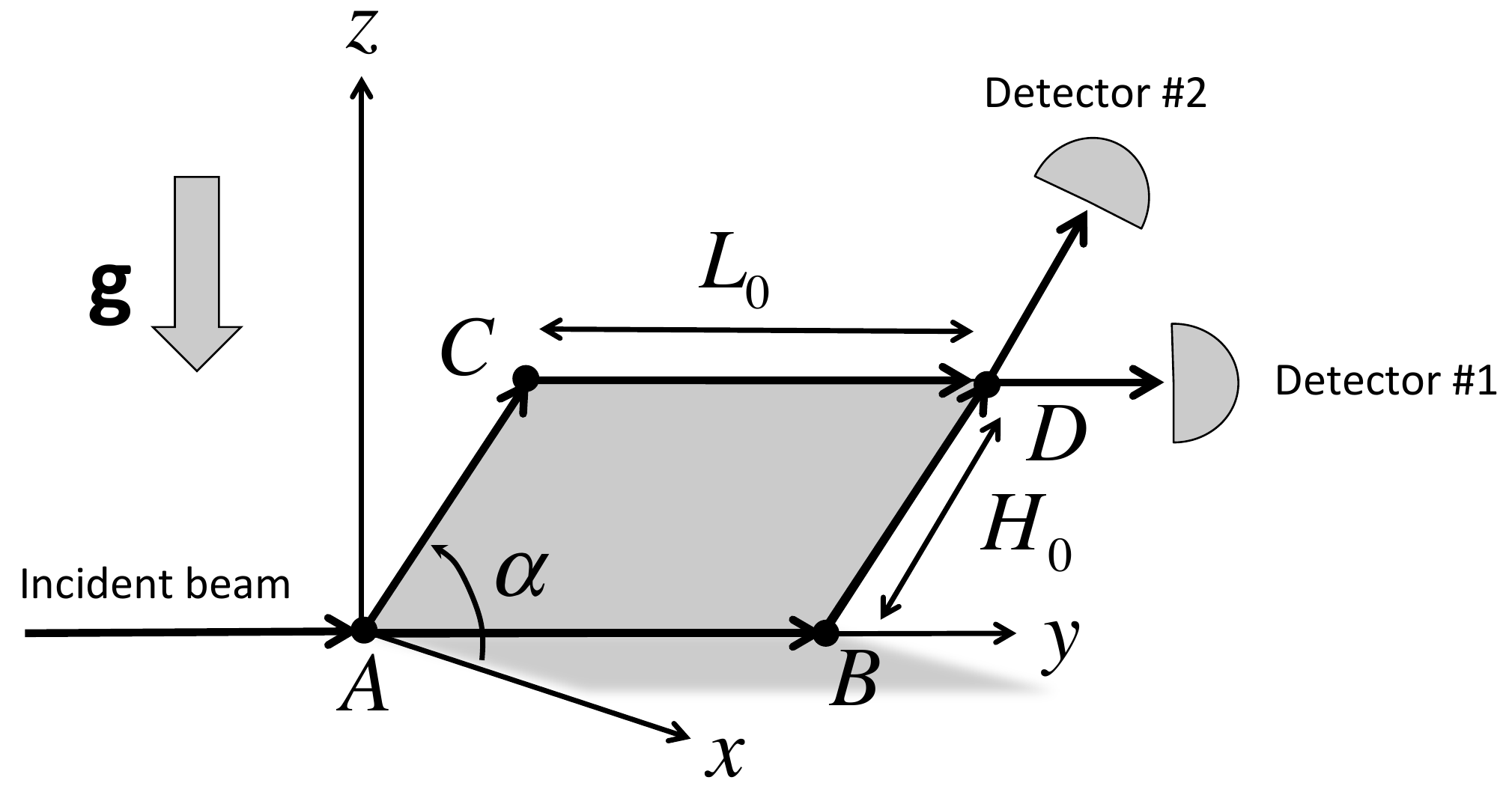}
\caption{A simplified diagram of the COW experiment where $\mathbf{g}$ represents the gravitational acceleration.}
\label{COW figure}
\end{center}
\end{figure}

The double-slit experiment is an example of a particle traveling paths of different lengths.  An example of interference experiment where the wave packet of the particle travels paths of equal length but in unequal times is the Colella-Overhauser-Werner (COW) experiment which observed the first quantum mechanical gravitational phase shift \cite{COW}.  As shown in the simplified diagram in  Fig.~\ref{COW figure}, in this experiment a perfect silicon crystal splits an incident beam of neutrons at $A$ into two beams which travel paths $ABD$ or $ACD$ until recombined at $D$.  From there, the particles will be directed into detector  \#1 or \#2.
When rotating the crystal along a horizontal axis (in our case, the $y$-axis), the phase shift arising from the fact that gravity reduces the particle's momentum traveling the leg $CD$ compared to $AB$ leads to a modulation of the detection probabilities in the two detectors.  Using the COW experiment with unstable particles to investigate the gravitational equivalence principle has been discussed recently by Bonder, et al. \cite{Bonder}, and that work motivated this section.

To obtain the phase shift for QUUPs in the COW  experimental setup shown in Fig.~\ref{COW figure}, we can use Eq.~(\ref{final Psi E}) while setting $V(\mathbf{r}) = mgz$, where $m$ is the mass of the particle and $g$ is the acceleration due to gravity.  
Let us assume that the beamsplitter at $A$ is ideal, so that the total wave function that is incident upon beamsplitter $D$ has the following form:
\begin{eqnarray}
\psi_{E}(D) & = & A_{0}\chi(\mathbf{r}_{A})e^{ip(H_{0}+L_{0})/\hbar}e^{-(H_{0}+L_{0})/2\ell_{0}} \nonumber \\
&& \mbox{} \times \left({\cal T}_{\rm BS}{\cal R}_{\rm M}e^{i\tilde{\phi}_{ABD}} + {\cal R}_{\rm BS}{\cal R}_{\rm M}e^{i\tilde{\phi}_{ACD}}\right),
\label{COW psi}
\end{eqnarray}
where $s = H_{0} + L_{0}$ for both beams.  ${\cal R}_{\rm M}$ is the phase factor acquired after a mirror reflection,  while ${\cal R}_{\rm BS}$ (${\cal T}_{\rm BS}$) is the amplitude for the particle to be reflected (transmitted) at a mirror or beamsplitter such that \cite{GC}
\begin{eqnarray}
|{\cal R}_{\rm M}|^{2} & = & 1, \label{RM}\\
|{\cal R}_{\rm BS}|^{2} + |{\cal T}_{\rm BS}|^{2} &=& 1, \label{RT2} \\
{\cal R}_{\rm BS}{\cal T}_{\rm BS}^{*} + {\cal R}_{\rm BS}^{*}{\cal T}_{\rm BS} & = & 0. \label{RT*}
\end{eqnarray}
 The complex phase angles for the different paths are determined using Eq.~(\ref{complex phase}):
\begin{eqnarray}
\tilde{\phi}_{ABD} & = & \tilde{\phi}_{AB} +  \tilde{\phi}_{BD} =  \tilde{\phi}_{BD},
\label{tilde phi ABD} \\
\tilde{\phi}_{ACD} & = & \tilde{\phi}_{AC} +  \tilde{\phi}_{CD},  \nonumber \\
&= &  \tilde{\phi}_{AC}  -  \frac{m^{2}gH_{0}L_{0}\sin\alpha}{\hbar p_{0}} \nonumber  \\
& & \mbox{} +i\left(\frac{m^{3}g\Gamma}{2p_{0}^{3}} + \frac{mg\Gamma}{2p_{0}c^{2}}\right)H_{0}L_{0}\sin\alpha, 
\label{tilde phi ACD} 
\end{eqnarray}
 where we have used $\tilde{\phi}_{AB} = 0$ since $z = 0$ along this path.  The phases along the vertical portions $BD$ and $AC$ are the same for both paths,
 \begin{eqnarray}
 \tilde{\phi}_{AC} =  \tilde{\phi}_{BD} & = & -\frac{m^{2}g}{2\hbar p_{0}}\left[1- i\left(\frac{m\hbar\Gamma}{2p_{0}^{2}} + \frac{\hbar\Gamma}{2mc^{2}}\right)\right]H_{0}^{2}\sin\alpha, \nonumber\\
 & = & -\frac{m^{2}gH_{0}^{2}}{2\hbar p_{0}}\sin^{2}\alpha + i\left(\frac{m^{3}gH_{0}^{2}\Gamma}{4p_{0}^{3}} 
 		+ \frac{mgH_{0}^{2}\Gamma}{4p_{0}c^{2}}\right)\sin\alpha.
\label{tilde phi AC}
 \end{eqnarray}
 The complex phase difference can then be written as
\begin{eqnarray}
 \Delta\tilde{\phi}(\alpha) & = &  \tilde{\phi}_{ACD}  - \tilde{\phi}_{ABD}, \nonumber \\
   & \equiv & \Delta\phi_{\rm COW}(\alpha)  + i \Delta\phi_{\rm UCOW}(\alpha), 
  \label{delta ACD ABD}
 \end{eqnarray}
where  the real portion,
 \begin{equation}
 \Delta\phi_{\rm COW}(\alpha) =  -\left(\frac{m^{2}gH_{0}L_{0}}{\hbar p_{0}}\right)\sin\alpha \equiv -q_{\rm COW}\sin\alpha,
 \end{equation}
 is the usual COW phase shift, while the imaginary component arises if the particle is unstable:
 \begin{equation}
 \Delta\phi_{\rm UCOW}(\alpha) =   \left(\frac{m^{3}g\Gamma H_{0}L_{0}}{2 p_{0}^{3}}\right)\sin\alpha \equiv q_{\rm UCOW}\sin\alpha.
 \label{Delta UCOW}
 \end{equation}

 Both the real and imaginary phase differences arise from the horizontal portions of the paths ($AB$ and $CD$) because the phase differences from the vertical paths ($AC$ and $BD$) are the same.  By energy conservation, the momentum for the particle taking the upper path $CD$ is less than lower path AB: $p_{CD}   \simeq p_{0} - (m^{2}gH_{0}/p_{0})\sin\alpha$.  This results in a relative shift in the de~Broglie wavelength between the particle traveling $AB$ and $CD$, which gives rise to the usual COW effect.  The smaller momentum along the upper path also reduces $\ell_{CD}$, the average distance the unstable particle taking the upper path travels before decaying, compared to the lower path survival distance $\ell_{AB} = \ell_{0}$:
 $\ell_{CD} = (p_{CD}/m\Gamma) \simeq  \ell_{0}[1 - (m^{2}gH_{0}/p_{0}^{2})\sin\alpha].$

In the experimental setup,  detectors \#1 and \#2 are set to detect the particles only if they have not decayed.  Then, using Eq.~(\ref{COW psi}),  the probability that detector \#1 detects the undecayed particles is given by \cite{Werner book}
\begin{eqnarray}
P_{D1}(\alpha) &=& \left|e^{ip(H_{0}+L_{0})/\hbar}e^{-(H_{0}+L_{0})/2\ell_{0}}\right. \nonumber \\
&& \left. \mbox{}\times
	\left({\cal T}_{\rm BS}{\cal R}_{\rm M}{\cal R}_{\rm BS}e^{i\tilde{\phi}_{ABD}} + 
	{\cal R}_{\rm BS}{\cal R}_{\rm M}{\cal T}_{\rm BS}e^{i\tilde{\phi}_{ACD}}\right)\right|^{2},
 \nonumber \\
& = &   e^{-(H_{0}+L_{0})/\ell_{0}}\nonumber \\
&& \mbox{} \times \left|{\cal T}_{\rm BS}{\cal R}_{\rm BS}e^{i\tilde{\phi}_{ABD}} + {\cal R}_{\rm BS}{\cal T}_{\rm BS}e^{i\tilde{\phi}_{ACD}}\right|^{2},
\label{P1a}
\end{eqnarray}
where Eq.~(\ref{RM}) has been used and we have set $A_{0}\chi(\mathbf{r}_{A}) = 1$.
To simplify subsequent calculations, let us assume ${\cal T}_{\rm BS} = 1/\sqrt{2}$ and ${\cal R}_{\rm BS} = i/\sqrt{2}$, in which case
\begin{eqnarray}
\lefteqn{\left|{\cal T}_{\rm BS}{\cal R}_{\rm BS}e^{i\tilde{\phi}_{ABD}} + {\cal R}_{\rm BS}{\cal T}_{\rm BS}e^{i\tilde{\phi}_{ACD}}\right|^{2} =  \frac{1}{4} \left|e^{i\tilde{\phi}_{ABD}} + e^{i\tilde{\phi}_{ACD}}\right|^{2}, } \nonumber \\
	& = & \frac{1}{4}\left[e^{i(\tilde{\phi}_{ABD}- \tilde{\phi}_{ABD}^{*})} + e^{i(\tilde{\phi}_{ACD} - \tilde{\phi}^{*}_{ACD})}\right] 
	\nonumber \\
	&& \mbox{} + \frac{1}{4}\left[e^{i(\tilde{\phi}_{ABD} - \tilde{\phi}^{*}_{ACD})}  + e^{-i(\tilde{\phi}^{*}_{ABD} - \tilde{\phi}_{ACD})}\right], \nonumber \\
	& = & \frac{1}{4}\left[e^{-2{\rm Im}(\tilde{\phi}_{ABD})} + e^{-2{\rm Im}(\tilde{\phi}_{ACD})}\right] 
\nonumber \\
	&& \mbox{} + \frac{1}{4}e^{-{\rm Im}(\tilde{\phi}_{ABD} + \tilde{\phi}_{ACD})}\left[e^{i{\rm Re}(\tilde{\phi}_{ABD} - \tilde{\phi}_{ACD})}  + e^{-i{\rm Re}(\tilde{\phi}_{ABD} - \tilde{\phi}_{ACD})}\right]. \nonumber \\
	\label{amplitude 1 a}
\end{eqnarray}
Using Eqs.(\ref{tilde phi ABD})--(\ref{tilde phi AC}),
\begin{eqnarray}
{\rm Im}(\tilde{\phi}_{ABD}) & \simeq & \left(\frac{m^{3}g\Gamma H_{0}L_{0}}{2p_{0}^{3}}\right) \sin\alpha\left(\frac{H_{0}}{2L_{0}}\right),\nonumber \\
& \simeq & q_{\rm UCOW}\sin\alpha\left(\frac{H_{0}}{2L_{0}}\right),
\label{Im ABD} \\
&& \nonumber \\
{\rm Im}(\tilde{\phi}_{ACD}) & \simeq & q_{\rm UCOW}\sin\alpha\left(1 + \frac{H_{0}}{2L_{0}}\right).
\label{Im ACD} 
\end{eqnarray}
Inserting Eqs.~(\ref{delta ACD ABD}), (\ref{Im ABD}) and (\ref{Im ACD})  into Eq.~(\ref{amplitude 1 a}) gives
\begin{eqnarray}
\lefteqn{\left|{\cal T}_{\rm BS}{\cal R}_{\rm BS}e^{i\tilde{\phi}_{ABD}} + {\cal R}_{\rm BS}{\cal T}_{\rm BS}e^{i\tilde{\phi}_{ACD}}\right|^{2}} \nonumber \\
	& = & \frac{1}{4}\exp\left[-2q_{\rm UCOW}\sin\alpha\left(1 + \frac{H_{0}}{2L_{0}}\right)\right] 
	\nonumber \\
	&& \mbox{} +  \frac{1}{4}\exp\left[-2q_{\rm UCOW}\sin\alpha\left(\frac{H_{0}}{2L_{0}}\right)\right] 
	\nonumber \\
	&& \mbox{}  + \frac{1}{2}\exp\left[-q_{\rm UCOW}\sin\alpha\left(1 + \frac{H_{0}}{L_{0}}\right)\right]\cos\left(q_{\rm COW}\sin\alpha\right). \nonumber \\
	\label{amplitude 1 b}
\end{eqnarray}
Substituting Eq.~(\ref{amplitude 1 b}) into Eq.~(\ref{P1a}) gives the probability that the particle is detected by detector \#1:
\begin{eqnarray}
P_{D1}(\alpha) &=&  \frac{1}{4}e^{-(H_{0}+L_{0})/\ell_{0}}
\nonumber \\
&& \mbox{} \times \left\{
\exp\left[-2q_{\rm UCOW}\sin\alpha\left(1 + \frac{H_{0}}{2L_{0}}\right)\right] \right.
	\nonumber \\
	&&  \mbox{} + \exp\left[-2q_{\rm UCOW}\sin\alpha\left(\frac{H_{0}}{2L_{0}}\right)\right] 
	\nonumber \\
	&&  \mbox{} 
	+ 2\exp\left[-q_{\rm UCOW}\sin\alpha\left(1 + \frac{H_{0}}{L_{0}}\right)\right]
	\nonumber \\
	&& \hspace{.5in} \mbox{} \times \cos\left(q_{\rm COW}\sin\alpha\right)\bigg\},
	\label{P1}
\end{eqnarray}
The probability that the particle reaches detector \#2 instead  can be found in a similar manner.  

As in the double-slit experiment, the intensity of QUUPs observed by detector \#1 is obtained by dividing Eq.~(\ref{P1}) by the sum of the probabilities that the particles reached the detector by paths $ABD$ and $ACD$ separately, which gives
\begin{equation}
I_{D1} = \frac{I_{0}}{2}\left[1 + {\rm sech}(q_{\rm UCOW}\sin\alpha)\cos\left(q_{\rm COW}\sin\alpha\right)\right],
\label{I COW}
\end{equation}
where $I_{0}$ is observed intensity when $\alpha = 0$.  The intensity reduces to the usual COW result for stable particles ($q_{\rm UCOW} = 0$).
The interference visibility for the COW experiment is then given by
 \begin{equation}
{\cal V}_{\rm COW}= {\rm sech}\left(q_{\rm UCOW}\sin\alpha\right),
\label{COW visibility}
 \end{equation}
while the corresponding path predictability analogous to Eq.~(\ref{P DS}) obtained from Eq.~(\ref{P1}) is 
\begin{equation}
{\cal P}_{\rm COW} = {\tanh}\left|q_{\rm UCOW}\sin\alpha\right|.
\label{COW predictability}
\end{equation}
When stable particles are used, ${\cal P}_{\rm COW} = 0$ since the particle is equally likely to reach the detector by either path.  However, for QUUPs, additional which-way information is available since the particle has a higher probability of reaching the detector via the lower (faster) path than the upper (slower) path.  Unfortunately, for neutrons in a typical COW experiment, the available which-way information is extremely small.  Using $H_{0} = L_{0} = 0.1$~m, $v_{0} = 2200$~m/s for thermal neutrons, one finds ${\cal P}_{\rm COW} \simeq q_{\rm UCOW} \simeq 5 \times 10^{-15}$, which compares with $q_{\rm COW} \simeq 700$ for the usual COW effect.

 Together, Eqs.~(\ref{COW visibility}) and (\ref{COW predictability}) satisfy Eq.~(\ref{P V}), and have the same general form as the results for the double-slit experiment.  In fact, the intensity observed by detector \#1  for the COW experiment, Eq.~(\ref{I COW}), can be expressed in the same form as the double-slit intensity, Eq.~(\ref{I DS}), if we generalize the meaning of $\Delta s$ to be the displacement between two wave packets starting simultaneously at $A$ when the first arrives at $D$.  For the double-slit experiment, $\Delta s$ is simply the difference in path lengths, while for the COW experiment, $\Delta s \simeq (m^{2}gH_{0}L_{0}/p_{0}^{2})\sin\alpha$. 
 
 \section{Discussion}
 
 Let us now examine the experimental issues in detecting these QUUP which-way effects.  Unstable particles (e.g., neutrons \cite{Werner book} and metastable atoms \cite{Vassen})  have been used in interference experiments for many years.  In nearly all these cases, the lifetimes of the particles were much longer than the duration of the experiment and so could be neglected.  An exception is Pfau, et al. \cite{Pfau}, who studied the effect of the decay of an atomic excited state on the atomic diffraction pattern.  In this case, the lifetime of the excited state of the He* atoms used was so short (100~ns) that all the excited atoms had  decayed by the time they reached the detector.  Since we are interested in the interference of {\em undecayed} particles, an appreciable number need to reach the detector without decaying, which means $\ell_{0} \sim L$, where $L$ is the path length traveled by the particle.  In an actual experiment, $\ell_{0} = \langle v_{0}\rangle\tau$, where  $\langle v_{0}\rangle$ is the average speed of the beam particles and $\tau$ is their average lifetime.   While any unstable particle (e.g, radioactive nucleus or subatomic particle) can be used, the most practical candidates are likely to be excited atoms since they possess a considerable range in  values of $\tau$, and their excitation process can be controlled.

  In order to maximize  the effect of the particle instability in a double-slit or COW experiment, we need to maximize the ratio $\Delta s/\ell_{0}$, where $\Delta s$ is the final packet separation.  Decreasing $\ell_{0}$ means using a particle with a shorter lifetime, which would reduce the overall signal.    Increasing $\Delta s$ may be challenging due to the finite longitudinal coherence length $L_{\rm coh}$  of real particle beams.  For a double-slit experiment using a beam of particles with a Gaussian distribution of wave numbers  characterized by  standard deviation $\sigma_{k} = 1/2\sigma_{0}$, where $\sigma_{0}$ is the initial standard deviation of the spatial wave packet, we show in the Appendix that the total visibility  takes the form
  \begin{equation}
 {\cal V}_{\rm tot} = {\cal V}_{\rm G}{\cal V}_{\rm DS} = e^{-(\Delta s)^{2}/8\sigma_{0}^{2}}{\rm sech}\left(\Delta s/2\ell_{0}\right),
 \end{equation}
 where ${\cal V}_{\rm G} = e^{-(\Delta s)^{2}/8\sigma_{0}^{2}}$ is the Gaussian beam visibility, and $L_{\rm coh} \sim \sigma_{0}$.  Thus, to maximize the effect of instability without being  significantly suppressed by the finite coherence length, the  final wave packet separation should satisfy $\ell_{0} \lesssim \Delta s \lesssim \sigma_{0}$.   Due to this condition, observing the which-way effects with QUUPs in  double-slit and COW-like experiments will be challenging.  Ideally one would like to modify the particle's decay rate in one of the interference paths to maximize the effect.  Fortunately, this is possible, in principle, if one uses atoms in excited states and uses appropriately tuned cavities \cite{Krause PLA}. 

\section{Conclusions}
 
In conclusion, we have extended the complementarity between which-way information and interference fringe visibility, Eq.(\ref{P V}), to interference with quantum undecayed unstable particles (QUUPs) in the double-slit and Colella-Overhauser-Werner (COW) experiments.   We have also derived a formalism which allows one to investigate interference with QUUPs in other types of potentials.  Finally, using wave packets in a double-slit experiment with QUUPs, we have shown in the Appendix how a time-dependent interference pattern leads to the time-independent result.

\section*{Acknowledgements}

We   thank Yuri Bonder, Hector Hern\'{a}ndez-Coronado, and Daniel Sudarsky for extensive discussions which provided  crucial insights that led to this work.  We also   thank Samuel Werner for illuminating conversations.

\appendix

\section{Double Slit Experiment with  Unstable Particle Gaussian Wave Packets (GWPs)}

To investigate quantum interference with non-relativistic unstable particles, one can use two different approaches.  The first, most obvious, is to apply an exponential time factor $e^{-\Gamma t/2}$ to all wave functions, where $\Gamma$ is the decay rate of the particle.  This seems to imply that we would obtain the same interference as for stable particles except for an overall time dependence $e^{-\Gamma t}$   \cite{Facchi,Takagi}.  The second approach  is time-independent, applicable to time-translation invariant situations such as a steady-beam of particles, which we used in Section~\ref{Double-Slit section}.  Using this approach, we found that the  interference patterns obtained for unstable particles differ from the corresponding interference for stable particles due to the additional which-way information available when one uses unstable particles.   

The suggestion that these two approaches appear to give different answers is only illusionary.  In this Appendix we will show that when one actually calculates the time-independent probability of particle detection for the double-slit experiment using the time-dependent wave packets, one obtains the same result found  using the steady-beam approach.  This is consistent with what is known for stable particles---interference of a steady-beam of particles is equivalent to the time-averaged interference of a beam of particles described by wave packets \cite{Adams,Dicke}.

\begin{figure}[tp]
\begin{center}
\includegraphics[height=2.2in]{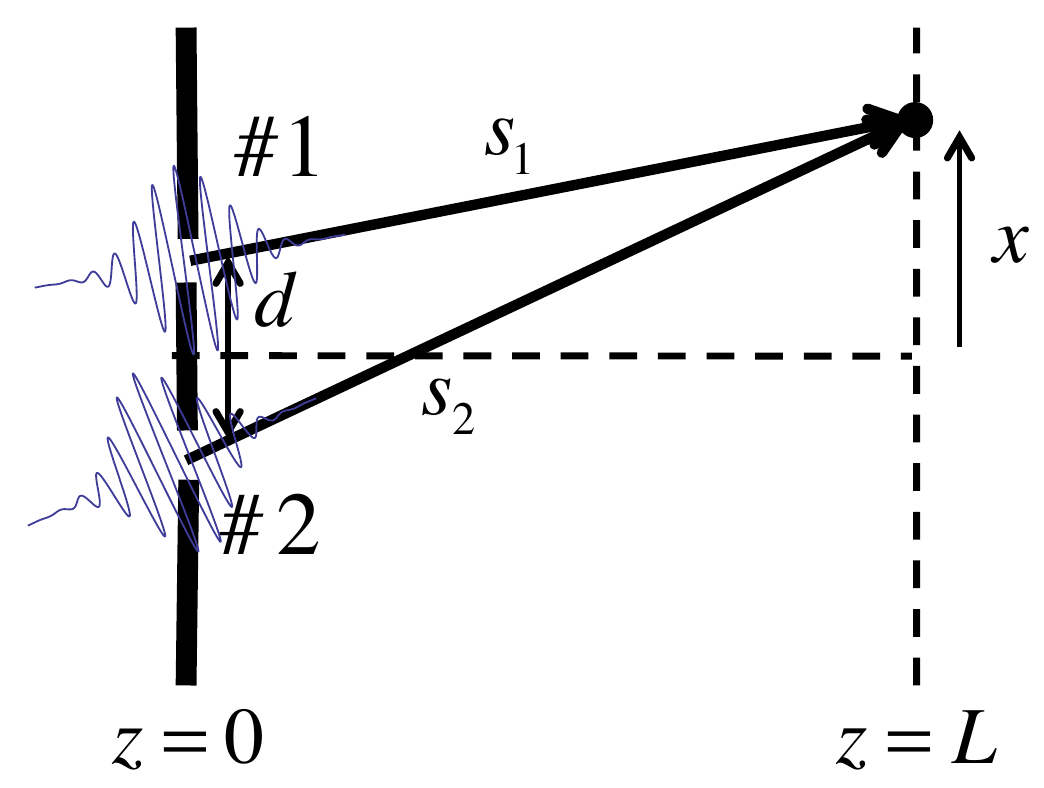}
\caption{Setup for double-slit experiment with Gaussian wave packets. Packets are shown at $t = 0$.}
\label{double-slit setup figure}
\end{center}
\end{figure}

The setup that we will use for the double-slit experiment is shown in Fig.~\ref{double-slit setup figure}.  Two slits of negligible width  located at $z = 0$ are separated by a distance $d$.
The particles are detected in the $xy$-plane at $z = L$.  Assuming the slits extend along the $y$-axis, the detection probability at $z = L$ will only depend on $x$.  We will assume that $x \ll L$ so that  the magnitude of the amplitude of the waves coming from the slits is the same.

To reach the detection point on the screen, the particle coming from the $i$th slit will have traveled the path of length $s_{i}$ as shown in Fig.~\ref{double-slit setup figure}.  For our simple treatment, we  will assume the unstable particles with decay rate $\Gamma$ are described by one-dimensional Gaussian wave packets (GWPs) traveling along these paths. (A more accurate calculation using fully 3-dimensional wave functions, as in Refs~\cite{Gondran,Viale, Beau}, is unnecessary for our purposes.) Specifically, the wave function for a GWP of width $\sigma_{0}$ centered at $s = s_{0}$ is given by
\begin{equation}
\Psi(s,t) \simeq \left(\frac{1}{2\pi \sigma_{0}^{2}}\right)^{1/4}e^{i[k_{0}s - \omega_{0}t)}e^{-(s-s_{0} - v_{g}t)^{2}/4\sigma_{0}^{2}}e^{-\Gamma t/2},
\label{GWP}
\end{equation}
where the packet group velocity is
\begin{equation}
v_{g} = \frac{hk_{0}}{m} = \frac{p_{0}}{m},
\end{equation}
where $k_{0}$ is the wave number, $\omega_{0} = \hbar k_{0}^{2}/2m$ is the angular frequency, $p_{0}$ is the particle's momentum, and $m$ is its mass.   To simplify calculations we are assuming that the time the particle takes to travel from the slits to the screen ($\Delta t = mL/p_{0}$) is much less than the time $t_{\rm spread} = 2m_{0}^{2}/\hbar$, for the wave packet to spread significantly, which implies
\begin{equation}
\sigma_{0} \gg \sqrt{\frac{\hbar L}{2p_{0}}}.
\end{equation}
[This condition is not strictly necessary since the broadening of a freely propagating wave packet does not change the interference pattern which depends on the (constant) longitudinal coherence length \cite{Klein}.] To compare results with the steady-beam approach, we will take the limit $\sigma_{0} \rightarrow \infty$.

We will assume that the packets will leave the slits in phase,  that the coordinate system is chosen such that $s = 0$ corresponds to the position of the detector, and that both packets are centered on their respective slits located at $s = -s_{i}$ at $t = 0$.  Then the wave packet coming from the $i$th slit will be written as
\begin{equation}
\Psi_{i}(s,t) \simeq \left(\frac{1}{2\pi \sigma_{0}^{2}}\right)^{1/4}e^{i[k_{0}(s+ s_{i})] - \omega_{0}t)}e^{-(s+s_{i} - v_{g}t)^{2}/4\sigma_{0}^{2}}e^{-\Gamma t/2}.
\label{ith GWP}
\end{equation}

Our goal is determine the probability that the unstable particle will be detected at the screen {\em irrespective of time.}  This will require a different procedure than simply finding the total wave function at $s = 0$  \cite{Viale}.  Instead, we need to find the total probability current $J(s,t)$ of the beams arriving at the detector.  Assuming the packets are released at $t = 0$, the probability of detection is then
\begin{equation}
P(s = 0) = \int^{\infty}_{0}dt\, J(s = 0, t).
\label{finding P}
\end{equation}

The next step involves finding the probability currents associated with each wave packet.  In one-dimension, the probability current for the $i$th packet alone is given by
\begin{eqnarray}
J_{i}(s,t) & = &  \frac{\hbar}{2mi}\left[\Psi_{i}^{*}(s,t)\frac{\partial\Psi_{i}(s,t)}{\partial s} -\Psi_{i}(s,t)\frac{\partial\Psi_{i}^{*}(s,t)}{\partial s} \right], \nonumber \\
& & \frac{\hbar}{m}{\rm Im}\left[\Psi_{i}^{*}(s,t)\frac{\partial\Psi_{i}(s,t)}{\partial s}\right],
\label{Ji}
\end{eqnarray}
where ``Im'' denotes the imaginary part of the argument.  The total probability current at $s = 0$ due to both packets is then
\begin{eqnarray}
J(s = 0,t) & = & \frac{N_{0}\hbar}{m}{\rm Im}\left\{\left[\Psi_{1}^{*}(0,t) + \Psi_{2}^{*}(0,t)\right]\left.\frac{\partial}{\partial s}\left[\Psi_{1}(s,t) + \Psi_{2}(s,t)\right]\right|_{s = 0}\right\}, \nonumber \\
		& = & N_{0}\left[J_{1}(0,t) + J_{2}(0,t) + J_{12}(0,t)\right],
\label{J0}
\end{eqnarray}
where 
\begin{equation}
J_{12}(0,t) =  \frac{\hbar}{m}{\rm Im}\left\{\Psi_{1}^{*}(0,t)\left[\frac{\partial\Psi_{2}(s,t)}{\partial s}\right]_{s = 0} + \Psi_{2}^{*}(0,t)\left[\frac{\partial\Psi_{1}(s,t)}{\partial s}\right]_{s = 0}\right\}.
\label{J12}
\end{equation}
A normalization constant $N_{0}$ has been included since the problem is not truly one-dimensional; there is obviously a non-zero probability of detecting the particle at other positions on the detection plane.
We can then use our GWPs given in Eq.~(\ref{ith GWP}) to calculate each of the parts of the total probability current at $s = 0$:
\begin{equation}
\Psi_{i}(0,t) =  \left(\frac{1}{2\pi \sigma_{0}^{2}}\right)^{1/4}e^{i(k_{0}s_{i} - \omega_{0}t)}e^{-(s_{i} - v_{g}t)^{2}/4\sigma_{0}^{2}}e^{-\Gamma t/2},
\label{ith PWP s = 0}
\end{equation}
and
\begin{equation}
\left.\frac{\partial\Psi_{i}(s,t)}{\partial s}\right|_{s = 0} = ik_{0}\Psi_{i}(0,t) - \left(\frac{s_{i} - v_{g}t}{2\sigma_{0}^{2}}\right)\Psi_{i}(0,t).
\label{derivative ith PWP}
\end{equation}
Substituting Eqs.~(\ref{ith PWP s = 0}) and (\ref{derivative ith PWP}) into Eq.~(\ref{Ji}) gives
\begin{eqnarray}
J_{i}(0,t) & = & \frac{\hbar k_{0}}{m}\left|\Psi_{i}(0,t)\right|^{2} = v_{g}\left|\Psi_{i}(0,t)\right|^{2}, \nonumber \\
&& \nonumber \\
		& = & v_{g} \left(\frac{1}{2\pi \sigma_{0}^{2}}\right)^{1/2}e^{-(s_{i} - v_{g}t)^{2}/2\sigma_{0}^{2}}e^{-\Gamma t}.
\label{final Ji}
\end{eqnarray}
Similarly, Eqs.~(\ref{ith PWP s = 0}) and (\ref{derivative ith PWP}) give us 
\begin{eqnarray}
\Psi_{1}^{*}(0,t)\left[\frac{\partial\Psi_{2}(s,t)}{\partial s}\right]_{s = 0} & = & ik_{0}\Psi_{1}^{*}(0,t)\Psi_{2}(0,t), \\
\Psi_{2}^{*}(0,t)\left[\frac{\partial\Psi_{1}(s,t)}{\partial s}\right]_{s = 0} & = & ik_{0}\Psi_{2}^{*}(0,t)\Psi_{1}(0,t),
\end{eqnarray}
so Eq.~(\ref{J12}) becomes
\begin{eqnarray}
J_{12}(0,t) & = &  2v_{g}\mbox{Re}\left[\Psi_{1}^{*}(0,t)\Psi_{2}(0,t)\right], \nonumber \\
	& = &  2v_{g} \left(\frac{1}{2\pi \sigma_{0}^{2}}\right)^{1/2}\cos[k_{0}(s_{1} - s_{2})]e^{-(s_{1} - v_{g}t)^{2}/4\sigma_{0}^{2}}e^{-(s_{2} - v_{g}t)^{2}/4\sigma_{0}^{2}}e^{-\Gamma t},
\label{final J12}
\end{eqnarray}
where ``Re'' denotes the real part of the argument, which is only non-zero when these packets overlap.  Thus, we obtain the following reasonable expression for the total probability current at the detector:
\begin{eqnarray}
J(0,t) & = &  N_{0}v_{g}\left\{\left|\Psi_{1}(0,t)\right|^{2} + \left|\Psi_{2}(0,t)\right|^{2} + 2\mbox{Re}\left[\Psi_{1}^{*}(0,t)\Psi_{2}(0,t)\right]\right\},
	\nonumber \\
	&& \nonumber \\
& = &  N_{0}v_{g}\left|\Psi_{1}(0,t) + \Psi_{2}(0,t)\right|^{2}.
\end{eqnarray}
We now have everything needed to calculate the detection probability.

To determine the probability of detecting the particle at position $x$, we first insert Eq.~(\ref{J0}) into Eq.~(\ref{finding P}) which gives
\begin{equation}
P = N_{0} \int^{\infty}_{0}dt\, \left[(J_{1}(0,t) + J_{2}(0,t) + J_{12}(0,t)\right] \equiv P_{1} + P_{2} + P_{12},
\label{P total}
\end{equation}
where $P_{1}$ and $P_{2}$ are the probabilities that the particle came from paths \#1 and \#2 if there was no interference, and $P_{12}$ is the interference term.  Using Eq.~(\ref{final Ji}),
\begin{eqnarray}
P_{i} & = & N_{0}\int^{\infty}_{0}dt\,J_{i}(0,t), \nonumber \\
& = &\frac{N_{0}v_{g}}{\sigma_{0}\sqrt{2\pi}}\int^{\infty}_{0}dt\,e^{-(s_{i} - v_{g}t)^{2}/2\sigma_{0}^{2}}e^{-\Gamma t}.
\label{Pi 1}
\end{eqnarray}
If $s_{i} \gg \sigma_{0}$, then the integrand is essentially zero at $t = 0$, and if there is a negligible probability that the particle will decay during the time it travels the width of the packet,
\begin{equation}
\frac{\Gamma \sigma_{0}}{v_{g}} \ll 1,
\end{equation}
we can safely replace the lower limit of integration by $-\infty$, giving
\begin{equation}
P_{i} \simeq  \frac{N_{0}v_{g}}{\sigma_{0}\sqrt{2\pi}}\int^{\infty}_{-\infty}dt\,e^{-(s_{i} - v_{g}t)^{2}/2\sigma_{0}^{2}}e^{-\Gamma t} \simeq N_{0}e^{-\Gamma s_{i}/v_{g}}.
\label{final Pi}
\end{equation}
This is just what we would expect classically.  The probability that an unstable particle traveling with speed $v_{g}$  reaches a distance $s_{i}$ during the travel time $t = s_{i}/v_{g}$ is $e^{-\Gamma t} = e^{-\Gamma s_{i}/v_{g}}$, which is (apart from the overall normalization constant $N_{0}$) just Eq.~(\ref{final Pi}).
Similarly, the interference contribution to the detection probability, $P_{12}$, is obtained using Eq.~(\ref{final J12}):
\begin{eqnarray}
P_{12} & = &  N_{0}\int^{\infty}_{0}dt\,J_{12}(0,t), \nonumber \\
		& \simeq &  \frac{2N_{0}v_{g}}{\sigma_{0} \sqrt{2\pi}} \cos[k_{0}(s_{1} - s_{2})] \int^{\infty}_{-\infty}dt\,e^{-(s_{1} - v_{g}t)^{2}/4\sigma_{0}^{2}}e^{-(s_{2} - v_{g}t)^{2}/4\sigma_{0}^{2}}e^{-\Gamma t}, \nonumber \\
		&& \nonumber \\
		& \simeq &2N_{0}\cos[k_{0}(s_{1} - s_{2})] e^{-(s_{1}-s_{2})^{2}/8\sigma_{0}^{2}}e^{-\Gamma(s_{1} +s_{2})/2v_{g}}.
\label{final P12}
\end{eqnarray}
Combining Eqs.~(\ref{P total}), (\ref{final Pi}), and (\ref{final P12}) then gives the total probability of being detected:
\begin{equation}
P = N_{0}\left\{e^{-\Gamma s_{1}/v_{g}} + e^{-\Gamma s_{2}/v_{g}} + 2e^{-(s_{1}-s_{2})^{2}/8\sigma_{0}^{2}}e^{-\Gamma(s_{1} +s_{2})/2v_{g}}\cos[k_{0}(s_{1} - s_{2})]\right\}.
\label{final P}
\end{equation}
For stable particles, Eq.~(\ref{final P}) reduces to
\begin{equation}
P = 2N_{0}\left\{1+ e^{-(s_{1}-s_{2})^{2}/8\sigma_{0}^{2}}\cos[k_{0}(s_{1} - s_{2})]\right\},
\label{final P stable}
\end{equation}
which is consistent with results of Adams, et al. \cite{Adams}. If the coherence length is very long, i.e., $\sigma_{0} \gg |s_{1} - s_{2}|$, then Eq.~(\ref{final P}) reduces to the steady-beam case derived in Section~\ref{Double-Slit section}:
\begin{equation}
P(\sigma_{0} \gg |s_{2} - s_{1}|) = N_{0}\left\{e^{-\Gamma s_{1}/v_{g}} + e^{-\Gamma s_{2}/v_{g}} + 
	2e^{-\Gamma (s_{1} + s_{2})/2v_{g}}\cos\left[k_{0}(s_{1}-s_{2})\right]\right\}.
\label{final P long}
\end{equation}
The only difference between Eqs.~(\ref{final P long}) and (\ref{P D}) is the normalization constant $N_{0}$ instead of $P_{0}$ which appears in the latter because the starting points of the wave packets are different. Thus, the steady-beam and wave packet approaches lead to the same detection probability for the double-slit experiment in the limit of long coherence lengths.  While it seems that one should get the stable particle interference pattern with QUUPs since the wave functions are the same apart from an overall  $e^{-\Gamma t}$ time-dependence, this does not take into account that the overlap of the packets depends on the difference in path lengths and the decay rate in a non-trivial way. 
Then when one integrates the instantaneous detection rate over time to obtain the total detection probability, the results for QUUPs differs from that of the corresponding stable particles.

\end{document}